\documentclass[aps,twocolumn,amsfonts,showpacs]{revtex4-1}

\pdfoutput=1
\usepackage{epsfig}
\usepackage{psfrag}
\usepackage{amsmath}
\usepackage{amssymb}
\usepackage{color}
\usepackage{appendix}
\usepackage{graphicx}
\usepackage{subcaption}
\usepackage{hyperref}
\usepackage{dcolumn}
\usepackage{bm}
\usepackage{float}
\usepackage{ragged2e}
\usepackage{microtype}

\begin{document}
\author{Jia-Hao Su$^1$}
\author{Chuan-Yin Xia$^1$}
\author{Wei-Can Yang$^{1,2}$}
\author{Hua-Bi Zeng$^1$}\email{ hbzeng@yzu.edu.cn}

\affiliation{$^1$ Center for Gravitation and Cosmology, College of Physical Science
and Technology, Yangzhou University, Yangzhou 225009, China}
\affiliation{$^2$ Department of Physics, Osaka Metropolitan University, 3-3-138 Sugimoto, 558-8585 Osaka, Japan}

\title{ Vortex-Antivortex Lattices in a Holographic Superconductor}
\begin{abstract}
We employ the Einstein-Abelian-Higgs theory to investigate the structure of vortex-antivortex lattices within a superconductor driven by spatial periodic magnetic fields. By adjusting the parameters of the external magnetic field, including the period ($\mathcal{T}$) and the amplitude ($B_0$), various distinct vortex states emerge. These states encompass the Wigner crystallization state, the vortex cluster state, and the suppressed state. Additionally, we present a comprehensive phase diagram to demarcate the specific regions where these structures emerge, contributing to our understanding of superconductivity in complex magnetic environments.

%When the magnetic strength is in the appropriate %interval, the vortex-Antivortex  pairs will form a %Wigner crystallization that exhibits rotational %symmetry. The vortices will form a cluster so the %vortex clusters and antivortex clusters will form %the Vortex-Antivortex cluster pairs that exhibit %mirror symmetry due to the property of Wigner %crystallization and periodic magnetic field. In %the cases of suitable period number, as the %magnetic strength increases, the superconductor %will show the Merssner state, vortex state %including Wigner crystallization morphology and v%ortex clusters morphology, suppressed state and %normal state in turn.

\end{abstract}

\maketitle
\section{introduction}

Type II superconductors are a distinct class of superconducting materials characterized by higher critical fields and stronger diamagnetism compared to Type I superconductors \cite{blatter1994vortices,de1966superconductivity,Ginzburg2009,rosenstein2010ginzburg,tinkham2004introduction}. The type of superconductor can be determined by the characteristic parameter $\kappa=\lambda/\xi$, which is defined as the ratio of the magnetic penetration depth ($\lambda$) to the coherence length of the order parameter ($\xi$). When $\kappa < \frac{1}{\sqrt{2}}$, the interface energy between the superconducting and normal states is positive, resulting in diamagnetism and classifying it as a Type I superconductor. On the other hand, when $\kappa > \frac{1}{\sqrt{2}}$, the interface energy is negative, resulting in the formation of the mixed state\cite{ms1,ms2,ms3,bogdanov1989thermodynamically,ms4,geurts2010vortex,shanenko2011extended,vagov2020universal}, i.e., the normal state occurs as vortices emerge on the superconductor to maintain a minimal free energy, which characterizes type II superconductor.                                                                               

Specifically, upon the application of a weak, uniform magnetic field to a superconductor, diamagnetic surface currents are induced inside the superconductor because of the presence of an external field outside the superconductor-vacuum interface. The penetration of the magnetic field within the superconductor follows the expression $B(x) \sim B_0 e^{-x/\lambda}$, where $B_0$ is the external magnetic field and $x$ is the direction perpendicular to the interface. Magnetic fields decay exponentially within the superconductor, which can be obtained from the London equation. This phenomenon is known as the Meissner effect. When the magnetic field increases to a critical point, the superconductor and magnetic field "meet a compromise" forming Abrikosov vortices, which can be arranged as triangular or square lattices\cite{abrikosov1957magnetic,gla2,gla3,gla4,gla5}.

Unlike the effects of a uniform magnetic field, the application of a periodic magnetic field to a superconductor gives rise to a more diverse range of phenomena, notably the emergence of vortex-antivortex pairs. These phenomena have been extensively studied in nonlinear G-L theory, yielding a wealth of insights \cite{milovsevic2003superconducting,milovsevic2004vortex,milovsevic2005vortex,geurts2006symmetric,berdiyorov2006novel,berdiyorov2009kinematic}. 
Nonetheless, the study of magnetic fields in strongly coupled superconductors is constrained by the limitations of the G-L equation. Therefore, the exploration of such systems necessitates the utilization of alternative theoretical tools and approaches. Recently, it became possible to address magnetic field problems in strongly coupled type II superconductor using the holographic duality discovered in string theory, mapping the condensed matter problem to a gravitational problem in one higher dimension \cite{witten1998anti,gubser1998gauge,maldacena1999large,gubser2008breaking,3h}. 
The charged scalar field represents the superconducting order parameter in dual quantum field theory, and its condensation leads to the formation of a superconducting phase in boundary theory. The scalar can have a nonzero profile with lower free energy than the trivial zero solution when the temperature of the black hole is low enough. Many previous studies on the effects of an external magnetic field\cite{albash2008holographic,albash2009phases,nakano2008critical,natsuume2022holographic} in the holographic superconductor model have laid the foundation for our work. The single vortex solution\cite{sv1,sv2,sv3,sv4}, vortex lattice solution\cite{vl1,vl2,xia2022holographic,baggioli2023sit} and periodic solution called holographic checkerboard\cite{withers2014holographic,donos2013competing} have already been obtained. Notably, all these previous studies with magnetic field have focused primarily on the effects of a uniform magnetic field, thus how the periodic magnetic field affects the holographic superconductivity is a problem worth exploring.

In this research, we conducted an in-depth analysis of vortex-antivortex lattices within a holographic superconducting film subjected to a periodic magnetized field. When the period number reaches a certain point, no vortex-antivortex lattices will form on the superconductor. In this case, the vortex and antivortex will annihilate, the magnetic field equals zero everywhere and the superconductor state exists for the arbitrary value of $B_0$, unless the superconducting state is completely disrupted by an extremely strong magnetic field. If the period number is kept within a suitable range, as the magnetic strength increases, the superconductor will go through the Merssner state, vortex state (including Wigner crystallization with rotational symmetry and vortex clusters with mirror symmetry), suppressed state, and eventually the normal state. It is worth noting that, in the context of our study, we also discovered that at a critical magnetic field strength, vortex-antivortex pairs begin to proliferate, forming larger clusters. This phenomenon leads to the emergence of long-range order within the system.

The paper is organized as follows. In Sec. \ref{H}, we introduce the holographic model and illustrate our setup, including the necessary theoretical framework and methodologies. In Sec. \ref{VAP}, we show various configurations of vortex-antivortex lattices in a superconducting film with an external periodic magnetic field and explore the dynamics of these lattices within the film. In Sec. \ref{P}, we show the phase diagram taking the period $\mathcal{T}$ and the strength of the external magnetic field $B_0$ as variables. This phase diagram provides a visual representation of the different phases and states exhibited by the superconducting film under varying conditions. We summarize our results in Sec. \ref{S}.

\section{Holographic model}\label{H}

In $3+1$ dimensional Anti-de Sitter space-time, we adopted a holographic superconducting model with a gauge field and a complex scalar field in the presence of a planar Schwarzschild black hole. This model is dual to a $2+1$ dimensional conformal field theory on the boundary. The action can be expressed as
\begin{equation}
S(\Psi,A_\mu)=\frac{1}{16\pi G_N} \int  d^4x \sqrt{-g}\Big[\mathcal{R}+\frac{6}{L^2}+\frac{1}{q^2}\mathcal{L}_{matter}(\Psi,A_\mu)
\Big]
\end{equation}
in which, a complex scalar field $\Psi$ is coupled to a U(1) gauge field $A_\mu$ in (3+1)D gravity with a cosmological constant related to the AdS radius as $\Lambda=-3/L^2$. The $G_N$ is the gravitational constant, The first two terms in parentheses are the gravitational part of the Lagrangian with the Ricci scalar $\mathcal{R}$, and the radius of the AdS spacetime $L$.
The Lagrangian matter field is
\begin{equation}
\mathcal{L}_{matter}(\Psi,A_\mu)=-\frac{1}{4}F_{\mu\nu}F^{\mu\nu}-|D_\mu\Psi|^2-m^2|\Psi|^2
\end{equation}   \label{lagrangian} where $F_{\mu\nu}=\partial_\mu A_{\nu}-\partial_\nu A_{\mu}$ is the component of the $U(1)$ gauge field and $\Psi$ is complex scalar field with mass $m$. $D_\mu$ is the covariant derivative written as
$D_\mu \Psi=\partial_\mu \Psi-iq_s A_\mu \Psi$ with the charge of the scalar field, as a Cooper pair, $q_s$ = 2e. We work in the probe limit by taking the $q \rightarrow \infty$, which means that the matter fields decouple from gravity, so standard Schwarzschild-AdS black brane to provide a constant temperature with metric.
\begin{equation}
    ds^2 = \frac{l^2}{z^2}\left(-f(z)dt^2 - 2dt\,dz + dx^2 + dy^2\right)
\end{equation}
$z=0$ represents the Ads boundary and $z=z_{h}$ is the horizon of black hole. Without loss of generality, we can set $z_{h}=L=1$, then $f(z)=1-z^3$, and the Hawking temperature can be written as $T=3/4\pi$. 

With the action one can obtain the equation of motion for the scalar field
\begin{equation}
D^\mu D_\mu \Psi-m_s^2 \Psi=0
\end{equation}
and the equation of motion for the vector field
\begin{equation}
\partial^\nu F_{\nu\mu}-iq_s(\Psi^*D_\mu\Psi-\Psi D_\mu\Psi^*)=0
\end{equation}
A trivial solution with $\Psi=0$ can be easily obtained from the equation mentioned above. However, when the potential on the AdS boundary is increased beyond a critical value, a non-zero solution for the scalar field appears. This indicates that for temperatures below $T_c$, a charged scalar operator has condensed, leading to the breaking of U(1) symmetry and the emergence of superconductivity.

The conformal dimensions of the scalar field can be obtained by setting $m_s^2=-2$, which yields $\Delta=\frac{3}{2}\pm \sqrt{\frac{9}{4}-m^2L^2}=\frac{3}{2} \pm \frac{1}{2}$. Therefore, the expansion of the scalar field solution near the AdS boundary takes the form of
\begin{equation}
\Psi = \phi z + \psi z^2 + \mathcal{O}(z^3)
\end{equation}
where $\phi|_{z=0}$ is set to be zero as a boundary condition when solving the model.
Similarly, the gauge field near the AdS boundary can be expressed as 
\begin{equation}
    A_\nu =a_\nu +b_\nu z + \mathcal{O}(z^2)
\end{equation}
where $a_t=\mu$ represents the chemical potential, if the value of $\mu$ exceeds a critical value $\mu_c=4.07$, the U(1) gauge symmetry of the system is spontaneously broken, resulting in a finite-valued solution for the expectation value of the scalar operator $\left \langle O \right \rangle =\psi |_{z=0}$. In the Eddington coordinate, the current $j_{\mu}$ is related to $b_{\mu}$ through $j_{\mu} = -b_{\mu} - \partial_{\mu} a_{t} + \partial_{t} a_{\mu}$ and the temperature T is related to $\mu$ through $T=(\mu_c/\mu)T_c$, following the holographic dictionary. In the case of a superconductor, the Neumann boundary condition for $A_x$ and $A_y$ is fixed as $j_x=j_y=0$ at $z=0$.

In order to apply an external periodic magnetic field to the superconducting film at $t=0$, we turn on 
\begin{align}
A_x=A_0sin(\mathcal{T}\frac{y\pi}{l})\\
A_y=-A_0sin(\mathcal{T}\frac{x\pi}{l})
\end{align}
where $\mathcal{T}$ represents the period number, $l$ represents the size of superconductor. The magnetic field can be obtained by taking the curvature of the vector potential
A using the equation $B=\nabla\times A$. 
In two dimensions, this equation can be expressed as 
\begin{equation}
B=-B_0[cos(\mathcal{T}\frac{x\pi}{l})+cos(\mathcal{T}\frac{y\pi}{l})]
\end{equation}
where $B_0=A_0\mathcal{T}\pi/l$.
A periodic magnetic field can be generated experimentally by arranging cubic magnetic dots in an array. By controlling the size, shape, and spacing of the magnetic dots, researchers can create a magnetic field that periodically varies in space. This allows researchers to study superconductor film directly in a controlled and repeatable way with periodic magnetic field\cite{exmt1,exmt2,del2016different}. 

For the numerical simulation, the Chebyshev spectral is used in the z-direction with 20 points. Fourier spectral is used in the (x,y) direction with $200\times200$ points. The evolution of time is simulated by the fourth-order Runge-Kutta method. The initial configuration at t=0 is chosen to be a superconducting state at fixed $l=50$ and $\mu=10$, so temperature $T=0.407T_c$.

%B=-B_0\mathcal{T}\frac{\pi}{l}[cos(\mathcal{T}\frac{x\pi}{l})+cos(\mathcal{T}\frac{y\pi}{l})]

\section{THE PATTERNS OF VORTEX-ANTIVORTEX LATTICES}\label{VAP}
This study investigates the consequences of introducing an external periodic magnetic field to a superconductor. This leads to the formation of various patterns of vortex-antivortex lattices. Similar to the experimental setup for generating vortices, we prepare a homogeneous superconducting state as the initial configuration, where the boundary condition is adjusted to introduce a periodic magnetic field onto the superconductor, resulting in system driving. To intuitively understand this, it can be considered as a regular cubic array of magnetic dots on the superconductor film. This cubic symmetry allows us to gain insight into the behavior of the system when exposed to external magnetic fields. The density of the array is correlated with the period number. As the period number increases, the array contains more magnetic dots, leading to a higher density. On the contrary, a smaller period number results in a lower density of magnetic dots in the array. The density of the array is a major factor in determining the system's overall behavior and characteristics, such as the magnetic interactions between the dots and their effect on the superconducting film. There are several physical effects that cause patterns to form, such as matching effects with ordered pinning arrays, which add to the pinning force due to the magnetic properties of the pinning centers. 

As $A_0$ increases, the process of a superconducting film can be described in four stages: the Meissner state, the vortex state including Wigner crystallization morphology and the vortex cluster morphology, the suppressed state, and the normal state. We concentrate on two types of type II superconductor patterns, namely vortex-antivortex pairs lattice patterns and suppressed patterns. We also investigate the progression of the phase transition between the vortex pairs state and the suppressed state, as well as the transition to the Merssner or normal state. The paper analyze superconductor instabilities under the magnetic field in a holographic model, and holographic periodic solution called checkerboard order has been obtained before, as well as the periodic lattice state of vortex-antivortex pairs can also be obtained in the G-L theory.

\begin{figure}[t]
    \centering
\includegraphics[trim=3.3cm 12.8cm 3.4cm 10.7cm, clip=true, scale=0.62, angle=0]{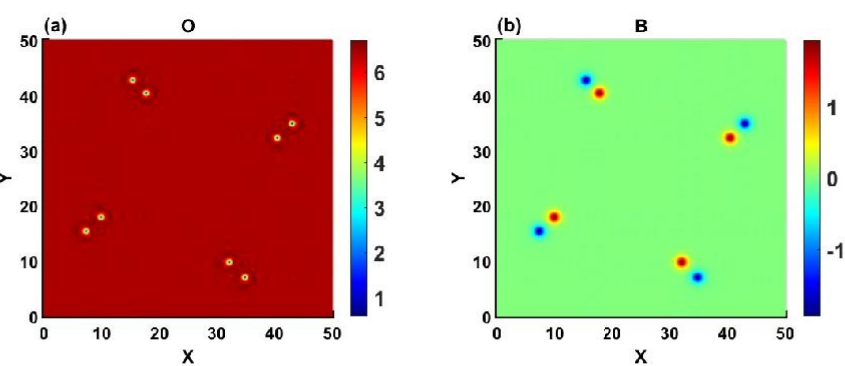}

\includegraphics[trim=3.3cm 12.8cm 3.4cm 10.7cm, clip=true, scale=0.62, angle=0]{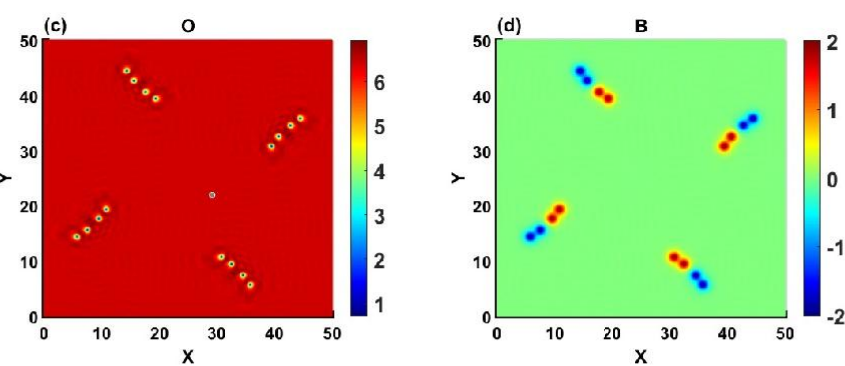}

\includegraphics[trim=3.3cm 12.8cm 3.4cm 10.7cm, clip=true, scale=0.62, angle=0]{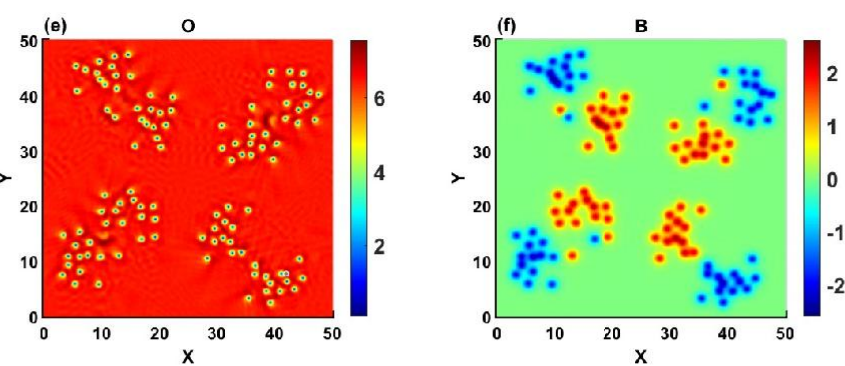}

\includegraphics[trim=3.3cm 12.8cm 3.4cm 10.7cm, clip=true, scale=0.62, angle=0]{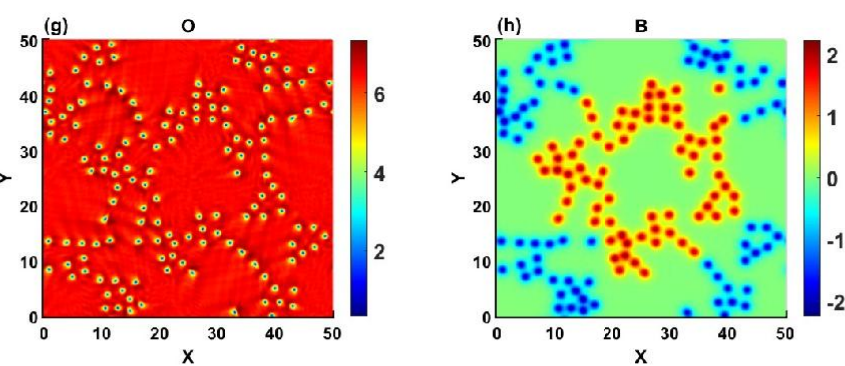}

\caption{\justifying Order parameter (left panels) and magnetic filed (right panels) when (a),(b) $B_0=1.0304$, (c),(d) $B_0=1.0556$, (e),(f) $B_0=1.3195$, and (g),(h) $B_0=2.9531$. The order parameter is defined as $\left \langle O \right \rangle =\partial_z\Psi$, the magnetic field is defined as $B=\partial_x A_y-\partial_y A_x$.
   In this situation, with a magnetic period $\mathcal{T}=2$, it can be considered as a primitive cell. The temperature $T=0.407T_c$. }.\label{fig1}
\end{figure}

\begin{figure}[H]
    \centering
\includegraphics[trim=3.3cm 12.8cm 3.4cm 10.7cm, clip=true, scale=0.62, angle=0]{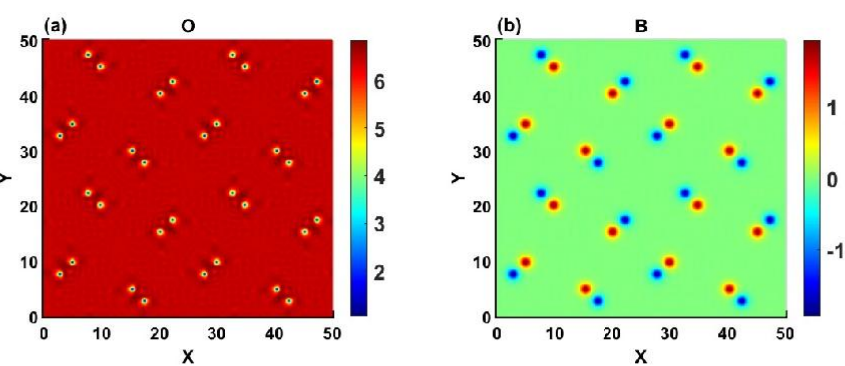}

\includegraphics[trim=3.3cm 12.8cm 3.4cm 10.7cm, clip=true, scale=0.62, angle=0]{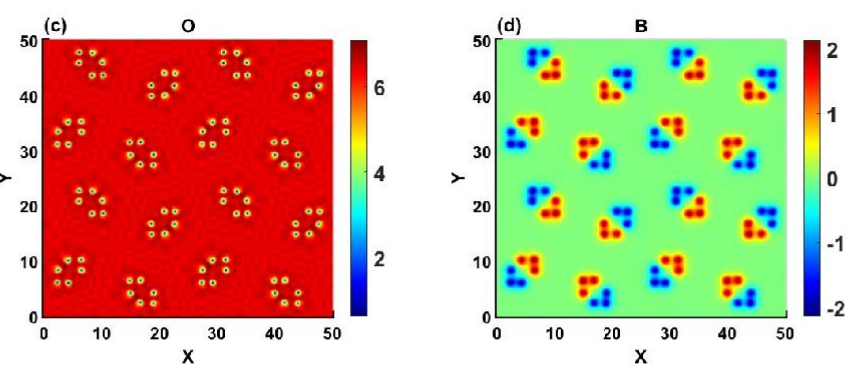}

\includegraphics[trim=3.3cm 12.8cm 3.4cm 10.7cm, clip=true, scale=0.62, angle=0]{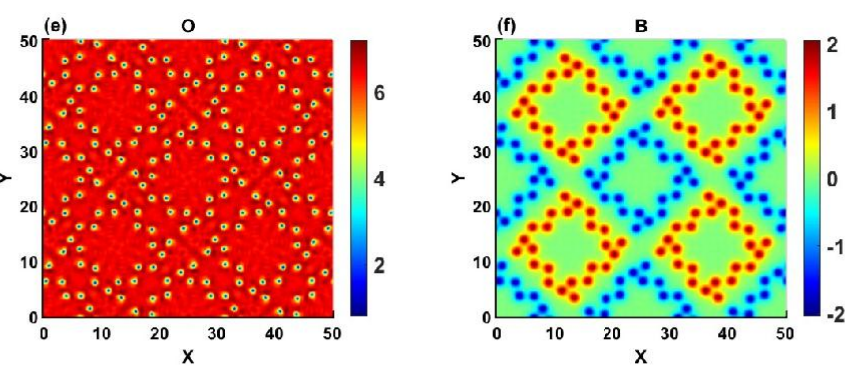}

\includegraphics[trim=3.3cm 12.8cm 3.4cm 10.7cm, clip=true, scale=0.62, angle=0]{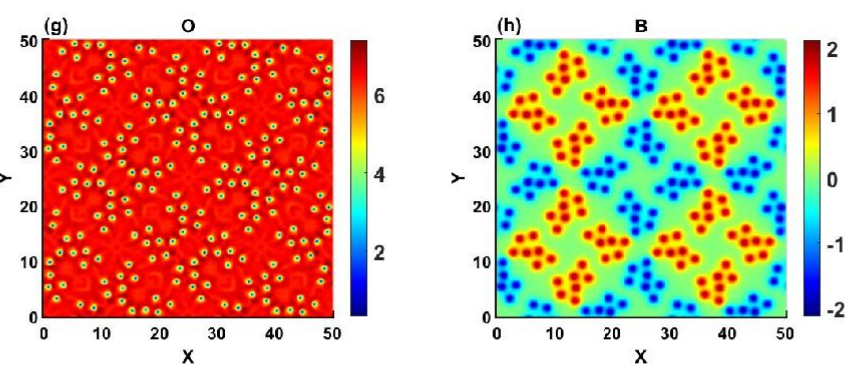}

\includegraphics[trim=3.3cm 12.8cm 3.4cm 10.7cm, clip=true, scale=0.62, angle=0]{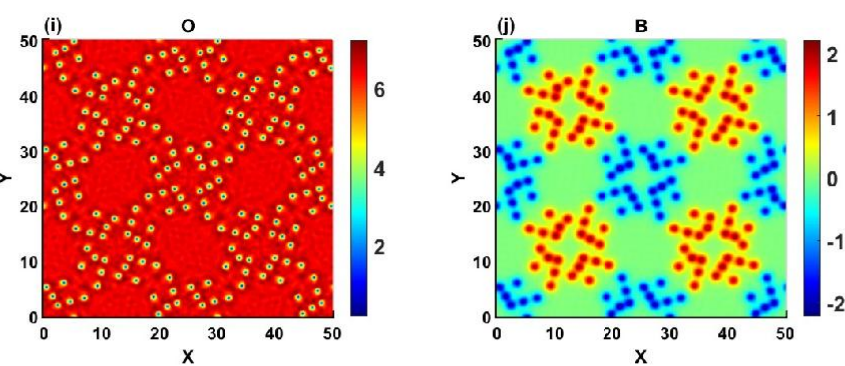}

\includegraphics[trim=3.3cm 12.8cm 3.4cm 10.7cm, clip=true, scale=0.62, angle=0]{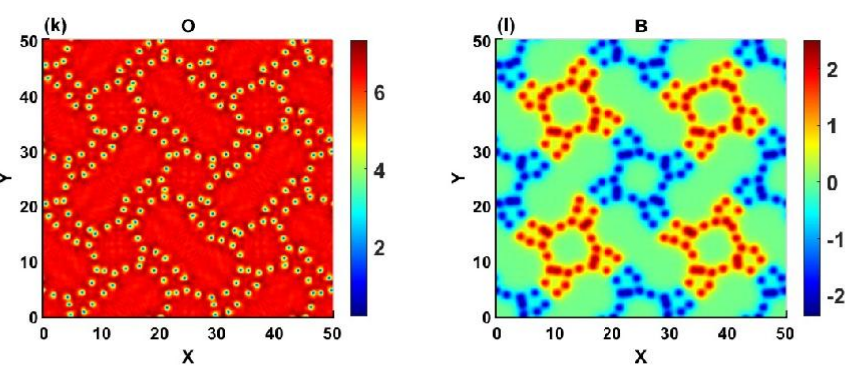}

\caption{Order parameter (left panels) and magnetic filed (right panels) when (a),(b) $B_0=2.0609$, (c),(d) $B_0=2.2117$, (e),(f) $B_0=3.5186$, (g),(h) $B_0=4.0212$, (i),(j) $B_0=6.2832$, and (k),(l) $B_0=8.5451$.
   The temperature $T=0.407T_c$ and the magnetic period $\mathcal{T}=4$}\label{fig2}
\end{figure}

\begin{figure}[h]
    \hspace{-0.5cm}
\includegraphics[trim=3.3cm 12.8cm 3.4cm 10.7cm, clip=true, scale=0.62, angle=0]{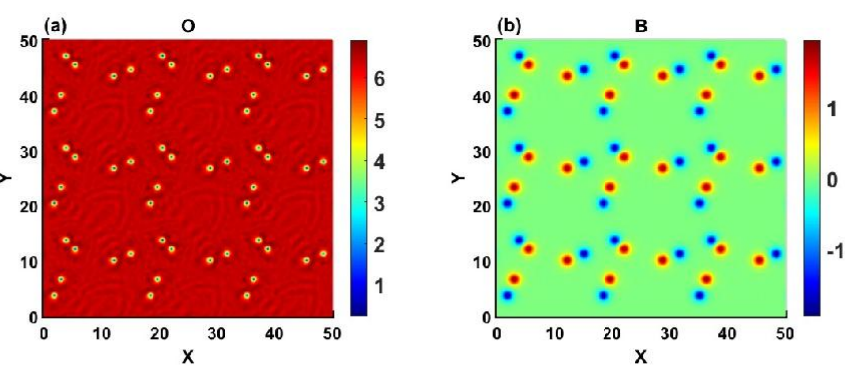}
 
 \hspace{-0.5cm}
\includegraphics[trim=3.3cm 12.8cm 3.4cm 10.7cm, clip=true, scale=0.62, angle=0]{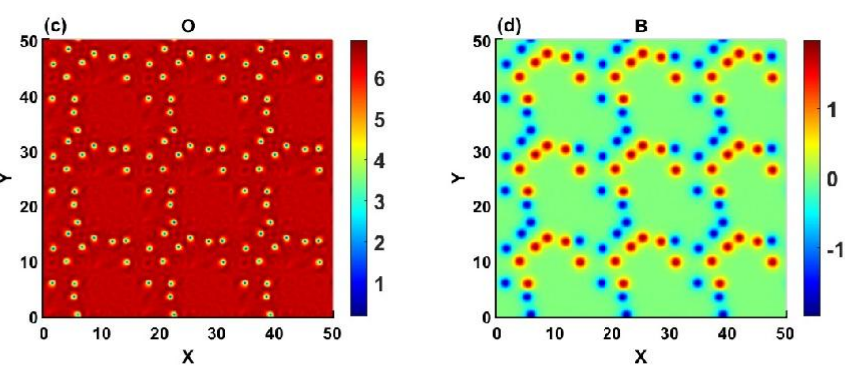}

\hspace{-0.5cm}
\includegraphics[trim=3.3cm 12.8cm 3.4cm 10.7cm, clip=true, scale=0.62, angle=0]{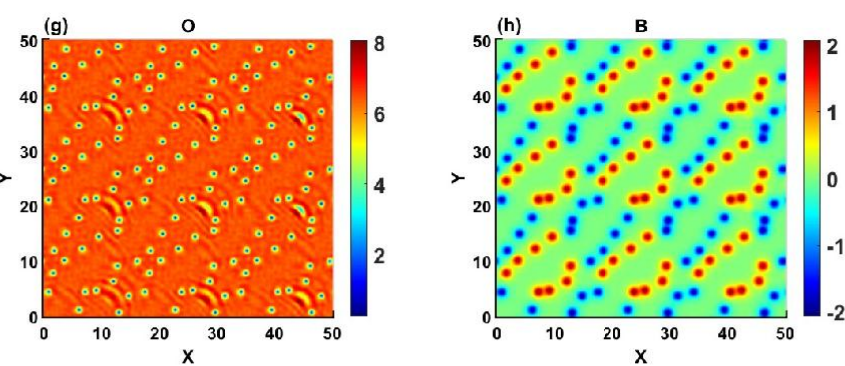}

\hspace{-0.5cm}
\includegraphics[trim=3.3cm 12.8cm 3.4cm 10.7cm, clip=true, scale=0.62, angle=0]{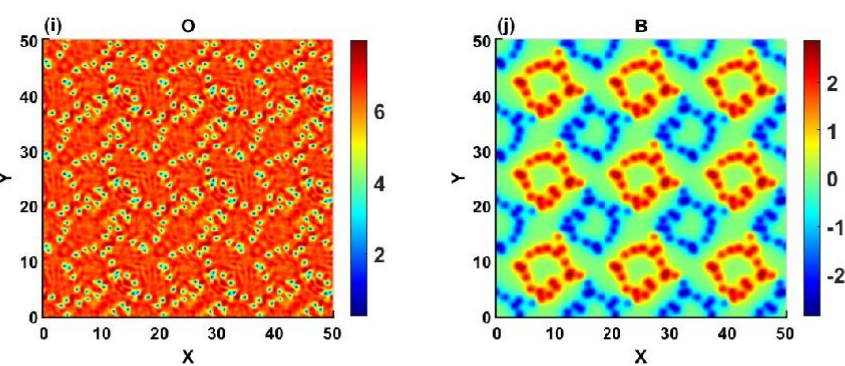}

\caption{\justifying Order parameter (left panels) and magnetic filed (right panels) when (a),(b) $B_0=3.7699$, (c),(d) $B_0=4.9009$, (e),(f) $B_0=5.2779$, (g),(h) $B_0=5.6549$, and (i),(j) $B_0=11.3097$.
   The temperature $T=0.407T_c$ and the magnetic period $\mathcal{T}=6$.}\label{fig3}
\end{figure}

\begin{figure}[h]
  \hspace{-0.5cm}
\includegraphics[trim=3.3cm 12.8cm 3.4cm 10.7cm, clip=true, scale=0.62, angle=0]{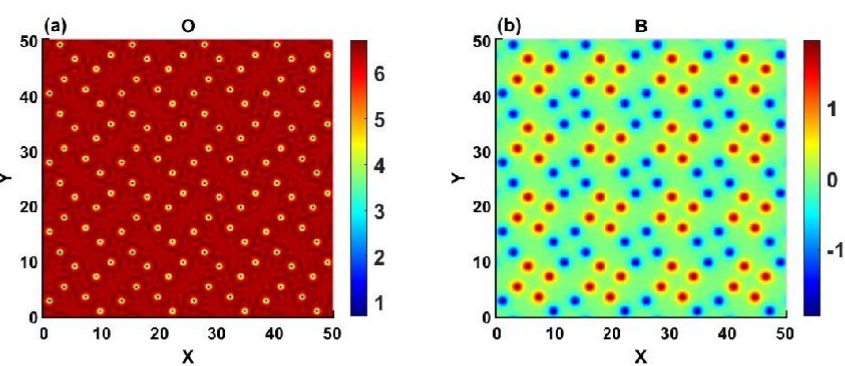}

\caption{\justifying Order parameter (left panels) and magnetic filed (right panels) when (a),(b) $B_0=7.0372$.
   The temperature $T=0.407T_c$ and the magnetic period $\mathcal{T}=8$}\label{fig4}
\end{figure}

%\subsection{the vortex state of superconducting film}\label{A}
In Fig. \ref{fig1}, \ref{fig2}, \ref{fig3} and \ref{fig4}, we show the vortex state patterns of the superconducting film, these patterns formed by the vortex-antivortex pairs exhibit various types of symmetry, including rotational symmetry, which can be expressed as an element of the group $SO(2)$ in the form of a matrix:\[
R(\pi/4) = \begin{bmatrix}
    0 & -1 \\
    1 & 0\\
\end{bmatrix}
\]
Additionally, the vortex cluster and antivortex cluster exhibit mirror symmetry form the Vortex-Antivortex cluster pair, which can be expressed as the equation $MX(x) = X(Mx)$, where M represents the mirror symmetry operation and X(x) represents the position of the vortex. Furthermore, the patterns display transitional symmetry in some situations, suggesting the Wigner crystallization property. In normal cases, the entire pattern can be obtained through symmetry operations. Starting with a vortex cluster, applying the mirror symmetry operation results in a Vortex-Antivortex cluster pair. By rotating this cluster pair, a primitive cell is obtained. Finally, by translating this primitive cell, the entire pattern can be generated.

We choose even period numbers to maintain a zero total flux passing through the superconductor film. This choice ensures that there is an equal number of vortex and antivortex, thereby preserving the symmetry in the patterns of vortex-antivortex pairs.
The patterns of $\mathcal{T}=2$ can be regarded as a single positive magnetic dot that acts in the center of the film, and thus can be treated as a primitive cell, specifically a $1\times1$ crystal lattice, as depicted in Fig.\ref{fig1}. Due to the positive magnetic dot being pinned to provide positive flux on the film center, when the magnetization reaches a sufficient level, the negative flux near the magnetic dot induces the magnetic field lines to "join" and form antivortices. Since the total flux must remain zero, the emergence of vortex-antivortex pairs occurs near the magnetic dot. When $\mathcal{T}=4$, as shown in Fig. \ref{fig2}, the patterns exhibit translation symmetry. Patterns can be identified as $2\times2$ crystal lattices. This situation can be interpreted as an array formed by four magnetic dots acting on the film. Around each magnetic dot, several vortex-antivortex pairs will form, resulting in the creation of vortex clusters, These vortex clusters represent a repeating units. The entire pattern observed in the superconducting film can always be obtained by translating a vortex cluster. Therefore, primitive cells can be associated with each magnetic dot. As the same way, the patterns of $\mathcal{T}=6$ shown in Fig.\ref{fig3} could be regarded as a $3\times3$ crystal lattice formed by an array of nine magnetic dots acting on the film.  Similarly, the patterns of $\mathcal{T}=8$ shown in Fig.\ref{fig4} could be regarded as a crystal lattice of $4\times4$ formed by an array of sixteen magnetic dots acting on the film. In each case, the patterns induce transitional symmetry, suggesting the Wigner crystallization property. Not that, when $\mathcal{T}=6$, the primitive cells do not exhibit rotational symmetry, so the entire pattern does not have rotational symmetry. The reason could be that the number of primitive cells is odd. In crystal lattices with an odd number of primitive cells, there is no equivalent rotation that can map one primitive cell onto another.

There is a special situation observed in Figs. \ref{fig1}(e)-(h), it is indeed observed that, when the system becomes very complex, the strict maintenance of symmetry becomes challenging due to the free evolution process of the system. When the system reaches a certain level of complexity, the interactions between the vortex-antivortex pairs become intricate. The presence of multiple factors, such as the density and arrangement of the pairs, their motion and dynamics, and the influence of thermal fluctuations, contribute to the intricate behavior observed in the system. The complex interplay of various factors leads to deviations from idealized symmetrical patterns. Understanding and characterizing the complexity that arises from the interactions of vortex-antivortex pairs and thermal fluctuations is a challenging task in superconductivity research\cite{epstein1981vortex}.

%The entire crystal structure can be generated by translating arbitrary primitive cells.

%\begin{figure}[H]

%\includegraphics[trim=4cm 0.1cm 1cm 0.1cm, clip=true, scale=0.24, angle=0]{3A80.jpg}
    
%\includegraphics[trim=4cm 0.1cm 1cm 0.1cm, clip=true, scale=0.24, angle=0]{3A90.jpg}

%\includegraphics[trim=4cm 0.1cm 1cm 0.1cm, clip=true, scale=0.24, angle=0]{3A110.jpg}

%\includegraphics[trim=4cm 0.1cm 1cm 0.1cm, clip=true, scale=0.24, angle=0]{3A150.jpg}

%\caption{Order parameter (left panels) and magnetic filed (right panels) when (a),(b) $B_0=8$, (c),(d) $B_0=9$, (e),(f) $B_0=11$, and (g),(h) $B_0=15$.
   %The temperature $T=0.406T_c$ and the magnetic period $\mathcal{T}=3$}.\label{fig2}
%\end{figure}

%So if the periodic magnetic field is considered as a square array of submicron cubic magnetic dots with perpendicular magnetization, the size of magnetic dots is related to magnetic-field strength$B_0$

%What is the reason that the system is more complex when $A_0\ge20$ for $\mathcal{T}=2$ than when $A_0\ge30$ for $\mathcal{T}=4,6,8$?

What is the cause of the complexity of the patterns increasing as the magnetic field strength increases when $\mathcal{T}=2$, but increasing initially and then stabilizing in a region when $\mathcal{T}=4,6,8$? The answer is that, when $\mathcal{T}=4,6,8$, there are multiple magnetic dots on the superconducting film, the vortex-antivortex pairs only can generate between the magnetic dots, the narrow area limits the complexity of the system as the vortex-antivortex pairs interact with each other. As the strength of the magnetic field increases, the area that can accommodate the vortex between the magnetic dots is restricted because the size of the area of influence of the magnetic dots also increases. So we can see the complexity of Figs. \ref{fig2} and \ref{fig3} rise first then down as increasing magnetic field, and the film is Meissner state when $\mathcal{T}=8$, $A_0=20$, i.e. $B_0=10.0531$. However, when $\mathcal{T}=2$, there is only one magnetic dot on the film, allowing the formation of multiple vortex-antivortex pairs. Sufficient area and strong magnetic strength induce the generation of large number pairs, that increases the complexity of the system. The system's complexity increases as more pairs are formed, creating intricate patterns. When $\mathcal{T}=4,6,8$, the magnetic dots are arranged in a homogeneous alignment on the film and vortex-antivortex pairs are generated in the gap. The size of the magnetic dots is fixed and directly linked to the magnetization, so the system reaches a point where the number of pairs begins to decrease as the magnetization increases, even though it is theoretically expected that a higher magnetization will lead to more pairs as a result of the more prominent fluctuations in the superconducting film. This can be explained as follows. As the size of the magnetic dots "expands" due to a stronger magnetization, the available space for the formation of pairs is limited. This restriction on pair formation keeps the system's complexity in check, even when the magnetic field strength is higher.

%Similar to the experimental setup for generating vortices, we prepare a homogeneous superconducting state as the initial configuration,
\begin{figure}[h]

   \hspace{-0.5cm}
\includegraphics[trim=3.3cm 12.8cm 3.13cm 10.7cm, clip=true, scale=0.62, angle=0]{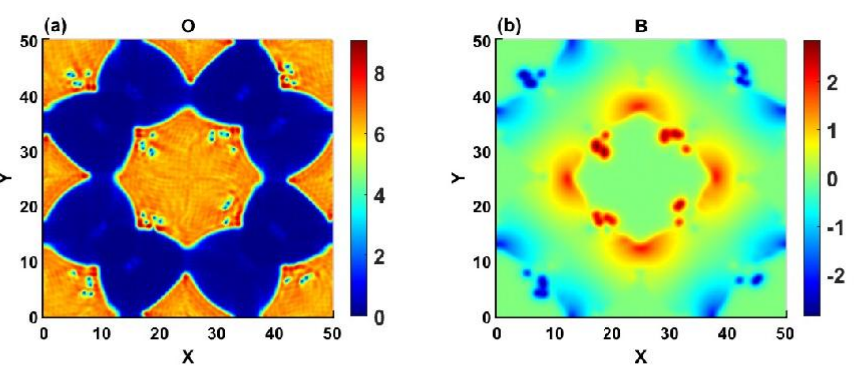}

\hspace{-0.5cm}
\includegraphics[trim=3.3cm 12.8cm 3.13cm 10.7cm, clip=true, scale=0.62, angle=0]{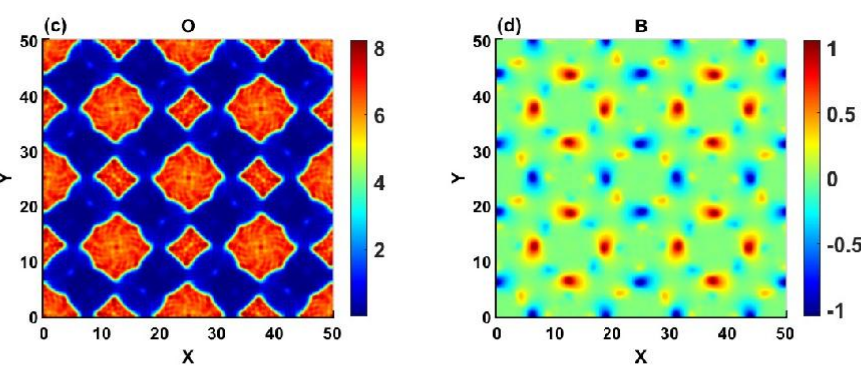}

\hspace{-0.5cm}
\includegraphics[trim=3.3cm 12.8cm 3.13cm 10.7cm, clip=true, scale=0.62, angle=0]{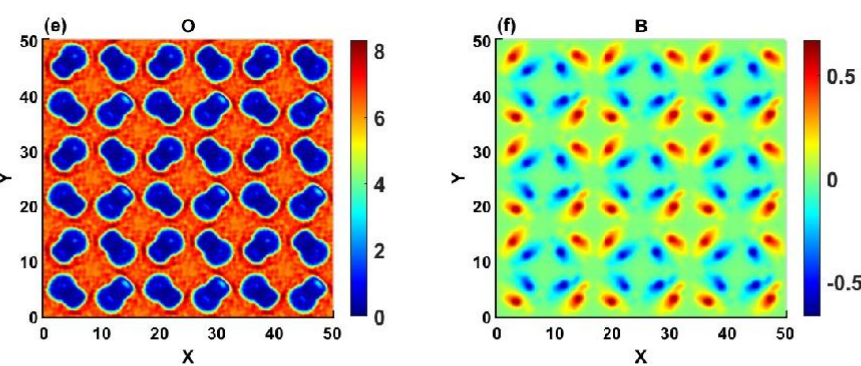}

\hspace{-0.5cm}
\includegraphics[trim=3.3cm 12.8cm 3.13cm 10.7cm, clip=true, scale=0.62, angle=0]{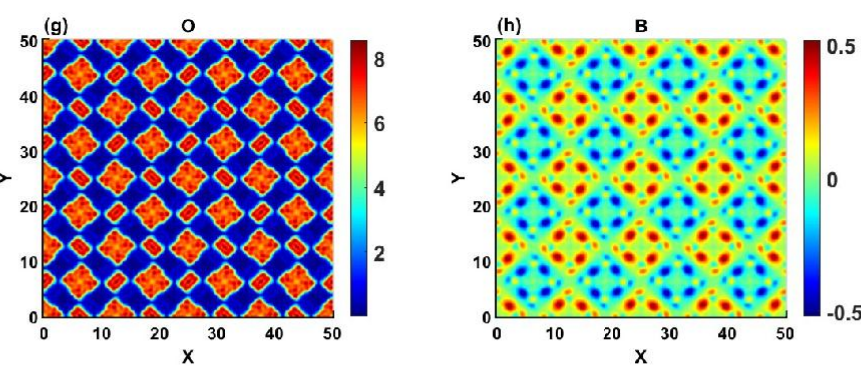}

\caption{\justifying Order parameter (left panels) and magnetic field (right panels) when (a),(b) magnetic period $\mathcal{T}=2$ and $B_0=4.7752$, (c),(d) magnetic period $\mathcal{T}=4$ and $B_0=9.5504$, (e),(f) magnetic period $\mathcal{T}=6$ and $B_0=14.3257$, and (g),(h) magnetic period $\mathcal{T}=8$ and $B_0=19.1009$.  
 The temperature $T=0.407T_c$ and the vector potential $A_0=38$.}\label{fig5}.
   
\end{figure}

\begin{figure}[h]
    \centering

\includegraphics[trim=3.3cm 12.8cm 3.4cm 10.7cm, clip=true, scale=0.62, angle=0]{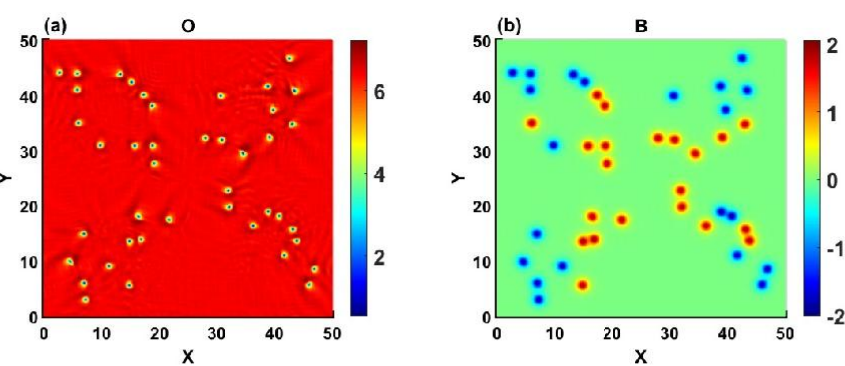}

\caption{\justifying Order parameter (left panels) and magnetic filed (right panels) when
   the temperature $T=0.406T_c$, the vector potential $A_0=38$, and magnetic field $B_0=4.7752$ and the magnetic period $\mathcal{T}=2$. Suppressed state back to vortex state. }\label{fig6}
\end{figure}

%\subsection{the suppressed state of superconducting film}\label{B}
When the strength of magnetic field  applied to the superconductor film reaches a certain threshold, a new state shown in Fig. \ref{fig5} emerges that the order parameter of the superconducting film is suppressed. This state is referred to as a metastable state because it can persist for a significant duration before transitioning back to either the vortex state or the superconducting state. The system returns to the vortex state shown in Fig. \ref{fig6}, it is a special case only when $\mathcal{T}=2$. And when $\mathcal{T}=4,6,8$, the system reverts to the Meissner state. These situations also can be explained by considering the periodic magnetic field as a square array of sub-micrometer cubic magnetic dots with perpendicular magnetization. As was mentioned above, as the magnetic field strength increases, the size of the influence of magnetic dots also grows, resulting in a smaller gap between them. Consequently, although the system has a tendency to generate vortex-antivortex pairs, the reduced gap between the magnetic dots causes the pairs to cluster closely together, resulting in annihilation when $\mathcal{T}=4,6,8$. This phenomenon only occurs when the magnetic field strength is relatively strong, and the specific properties of the magnetic dots lead to a cancellation of the magnetic field across the entire superconducting film. An example of the dynamic process of this phenomenon is shown in a movie that is given in the Supplementary Material\cite{movie1}. For $\mathcal{T}=2$ and $A_0=38$, i.e. $B_0=4.7752$ shown in Fig. \ref{fig6}, there is only one magnetic dot acting on the film; pairs will not be squeezed into a narrow gap between magnetic dots, so the vortex state is kept. An example of the dynamic process of this phenomenon is shown in a movie which is given in the Supplementary Material\cite{movie2}.

\section{THE PHASE DIAGRAM}\label{P}

\begin{figure}[h]
    \centering

\includegraphics[trim=3.3cm 9.4cm 3.4cm 9.55cm, clip=true, scale=0.62, angle=0]{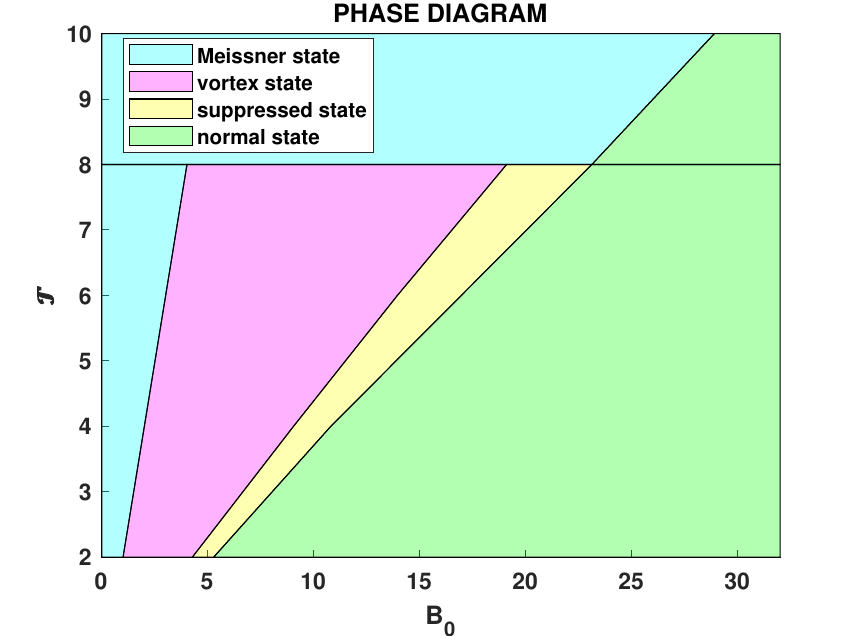}

\caption{
\justifying
The phase diagram of the holographic superconductor model with the magnetic period number ($\mathcal{T}$) on the x-axis and the strength of the magnetic field ($B_0$) on the y-axis. The cyan zone represents the meissner state in which magnetic field can 
not enter into superconductor. The magenta zone represents vortex state. The yellow zone represents suppressed state in which magnetic field suppressed order parameter for a long time firstly, then superconductor go back to vortex state or Meissner state. And the green zone represents normal state in which superconductor is completely destroyed by magnetic field.} \label{fig7}
\end{figure}

\justifying
We already show that type II superconductors exhibit different types of phases under a periodic magnetic field, including the Meissner state, the vortex lattice state, the suppressed superconductivity state, and the normal state in the rough phase diagram below. On the basis of the phase diagram, the basic physical landscape could be described. When $\mathcal{T}<8$, magnetization is too week in the cyan zone, so the magnetic field cannot enter the superconductor, as the magnetization gets stronger, the superconductor enters the vortex state and the quantity of vortex increases as $B_0$ increases. When the magnetic field strength increases to a critical value between the cyan zone and the magenta zone for a fixed $\mathcal{T}$, the superconductor transitions to the suppressed state, where the order parameter is suppressed due to the influence of a strong magnetic field on the superconducting film. Note that as $\mathcal{T}$ increases, the slope of the dividing line between different phases increases. To comprehend this behaviour, we use a helpful analogy of the periodic magnetic field acting on the superconductor film as a regular array of cubic magnetic dots acting on the film. As the magnetic period number $\mathcal{T}$ increases, the density of these magnetic dot arrays also increases on the film. The increased density of magnetic dot array provides additional stability to the vortex state, making it less susceptible to suppression by the external magnetic field. The emerge of suppressed state and it's evolution progress is interesting, the phenomenon and the progress including rich dynamic procedure between order parameter and magnetic field. Then as $B_0$ continue increases to a critical value between yellow zone and green zone for a fixed $\mathcal{T}$, the superconductor is completely destroyed by magnetic field. The situation of $\mathcal{T}>8$ is special, the density of the magnetic dot array is too high, resulting in a gap between two dots that is too small. This causes vortex annihilation before the system can become stable, leaving only superconducting and normal states.

\section{SUMMARY}\label{S}

In this paper, we have presented a holographic superconductor model for the numerical investigation of the behavior of a superconducting film in response to a periodic magnetic field. Due to the back reaction of black hole is not considered, the dynamic equations was derived from the Lagrangian matter field $\mathcal{L}_{matter}(\Psi,A_\mu)$, the temperature and external magnetic field of the superconducting system were regulated by setting boundary conditions for the dynamic equations. The results with the different period number ($\mathcal{T}$) and the strength of the magnetic field ($B_0$) show various structures of vortex states including the Wigner crystallization state, the vortex cluster state, and the suppressed state.
As the magnetic field increases, the behavior of the superconductor varies depending on the period number ($\mathcal{T}$). For $2< \mathcal{T} <8$, the superconductor will go through the Merssner state, the vortex state (including the Wigner crystallization and vortex clusters), the suppressed state and eventually the normal state. If $ \mathcal{T} >8$, the superconductor will transition directly from the Merssner state to the normal state. We also provided a phase diagram with the period number ($\mathcal{T}$) and the strength of the magnetic field ($B_0$) to sum up the results. Finally, the observation can be verified since the experimental developments enabled to produce a periodic magnetic field  with the magnetic cubic dot array on the top of superconductor directly\cite{exmt1,exmt2,del2016different}.

%According to the holographic dual dictionary, the temperature is regulated by the gauge field $A_t$, and the external magnetic field is regulated by the gauge fields $A_x$ and $A_y$. We given a constant $A_t$ and constructed a periodic magnetic field by imposing the periodic gauge fields $A_x$ and $A_y$, then the dynamic equations can be solved in the Eddington coordinate.

%We have used a numerical method based on a holographic superconductor model to investigate the behavior of a superconducting film driven by the periodic magnetic field.

\begin{acknowledgments}

\justifying
This work is supported by the National Natural Science Foundation of China (under Grants No. 11275233) and the Postgraduate Research \& Practice Innovation Program Jiangsu Province(KYCX22\_3450).
\end{acknowledgments}

\vspace{0mm}
\bibliography{Ref}
\end{document}